\documentclass{kapproc} 

\usepackage{procps} 

\usepackage[dvips]{graphicx}

\upperandlowercase

 \let\footnote\savefootnote

\normallatexbib 

\begin{document}

\articletitle
{Bubbles and Superbubbles}

\chaptitlerunninghead{Bubbles} 

\author{You-Hua Chu,\altaffilmark{1} Mart\'{\i}n A.\ 
 Guerrero,\altaffilmark{1,2} \& Robert A.\ Gruendl\altaffilmark{1}}

\affil{\altaffilmark{1}University of Illinois at Urbana-Champaign \
\altaffilmark{2}Instituto de Astrof\'\i sica de Andaluc\'\i a (CSIC)} 

\begin{abstract}
An isolated massive star can blow a bubble, while a group of massive
stars can blow superbubbles.  In this paper, we examine three 
intriguing questions regarding bubbles and superbubbles:
(1) why don't we see interstellar bubbles around every O star?
(2) how hot are the bubble interiors?  and
(3) what is going on at the hot/cold gas interface in a bubble?
\end{abstract}

\section{Introduction: Definition and Basic Facts}
Massive stars inject mechanical energy into the ambient medium
via fast stellar winds during their lifetime and supernova ejecta
at the end of their evolution.
The stellar winds and supernova ejecta sweep up and compress 
the ambient medium into shells, called {\it bubbles} or 
{\it superbubbles} depending on whether the energy source is an 
isolated massive star or a group of massive stars such as OB 
associations and young clusters.
As a bubble is blown by a single massive star, its size grows
up to a few $\times10$ pc before the star explodes, while 
a superbubble blown by a populous cluster or OB association 
can grow up to a few $\times10^2$ pc before exhausting all
massive stars.

The formation of a bubble is intimately dependent on the evolution 
and mass loss history of the central massive star.
Massive stars evolve from the main sequence (MS), through a 
luminous blue variable (LBV) or red supergiant (RSG) phase,
to the Wolf-Rayet (WR) phase before rushing toward the final
supernova explosion.
Along these evolutionary stages, a massive star loses mass
via tenuous fast (1000-2500 km~s$^{-1}$) stellar wind during 
the MS stage, copious slow (10-50 km~s$^{-1}$) wind during the 
RSG stage, copious slow or not-so-slow wind at the LBV stage, 
and fast stellar wind again during the WR stage.
A MS O-type star is most likely surrounded by the relics of its
natal cloud, thus its wind-blown bubble contains interstellar
material and is an {\it interstellar bubble}.
A WR star, on the other hand, is surrounded by the slow wind 
ejected by its progenitor during the RSG or LBV phase; thus 
its wind-blown bubble contains stellar material and is a
{\it circumstellar bubble}.

The most realistic hydrodynamic modeling of bubbles has been 
carried out by Garc\'{\i}a-Segura et al.\ (1996a, 1996b).
They take into account the stellar evolution and the mass loss 
history, forming an interstellar bubble around a MS O star at 
first and a circumstellar bubble around a WR star at the end.
Their models and the most well cited interstellar bubble model
by Weaver et al.\ (1977) both assume that the fast stellar wind
is shocked and forms a contact discontinuity with the dense
swept-up shell, and that the thermal pressure of the shocked
fast wind drives the expansion of the bubble.

The physical structure of a superbubble can be similar to that
of an interstellar bubble, if the fast stellar winds and the 
supernova ejecta are thermalized and confined in the superbubble
interior (Mac Low \& McCray 1988).
If supernovae occur near a superbubble shell, the impact on the
shell will produce signatures similar to a supernova remnant (SNR), 
and the shell will appear as a SNR-superbubble hybrid (Chu \& 
Mac Low 1990).

Finally, we note that planetary nebulae (PNe) are also bubbles,
as they are formed by the current fast wind of the central star
plowing into the circumstellar material shed by the progenitor 
via copious slow wind during the asymptotic giant branch (AGB)
and post-AGB stages, much like the formation of a WR bubble.
The optical morphology of a PN is frequently complicated by the 
presence of jets and collimated outflows, but the overall bubble
structure is still well shown in X-rays, and comparisons between
PNe and WR bubbles may help us understand both objects.

\section{Three Questions: Physical Structure of Bubbles}

The different layers of a wind-blown bubble can be observed at
different wavelengths: the photoionized swept-up shell is best
seen in the H$\alpha$ line, the shocked stellar wind can be 
detected in X-rays, and the interface is best observed in UV
or FUV.
Three questions naturally come to mind: \\ {\sl 
\indent\indent (1) Why don't we see interstellar bubble around every MS 
            O star? \\
\indent\indent (2) How hot is the bubble interior? \\
\indent\indent (3) What is going on at the hot/cold gas interfaces in a
            bubble?} \\
Below we present recent observations to answer these questions.

\begin{figure}
\centerline{\includegraphics[width=0.8\textwidth]{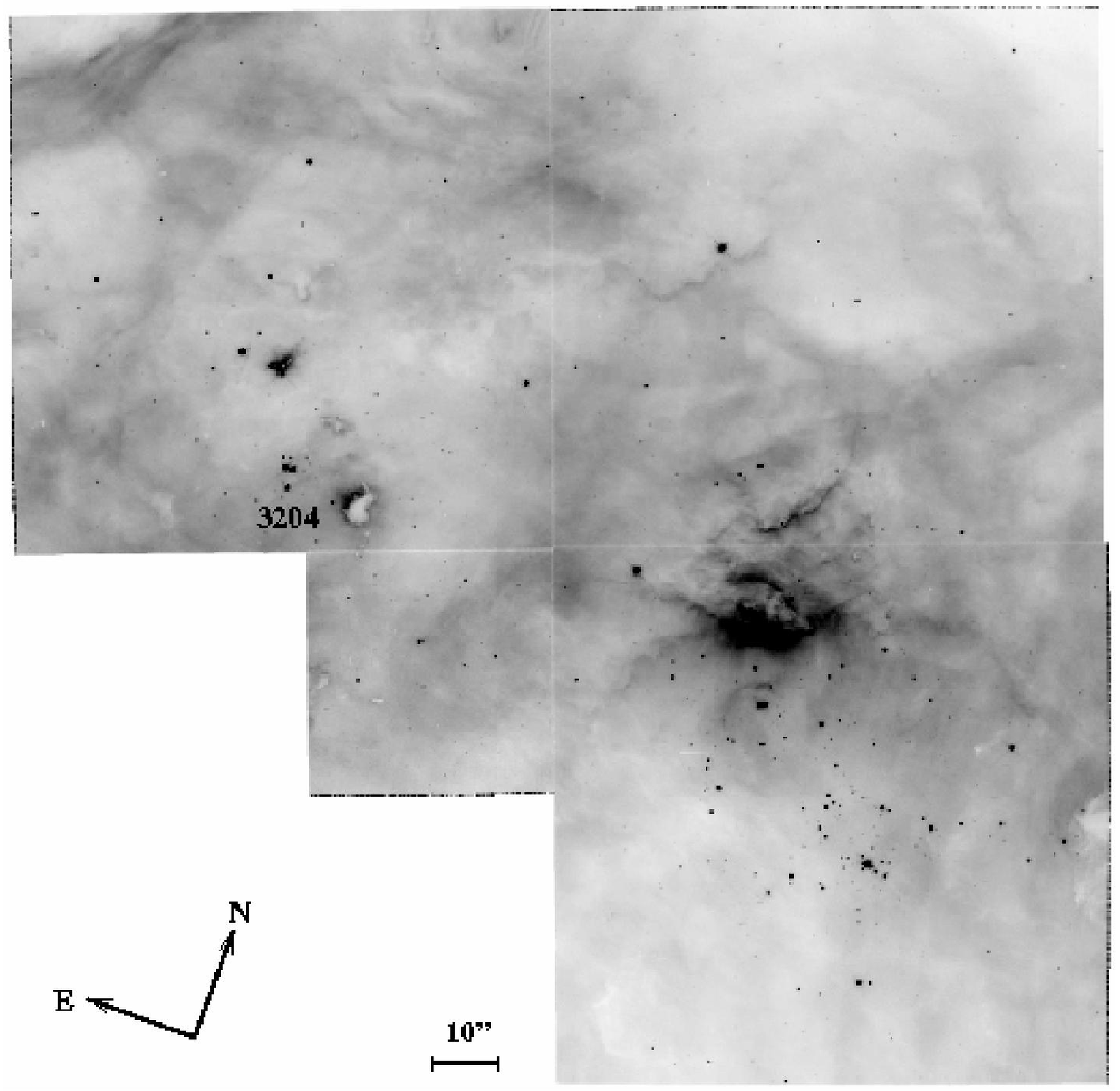}}
\centerline{\includegraphics[width=0.6\textwidth]{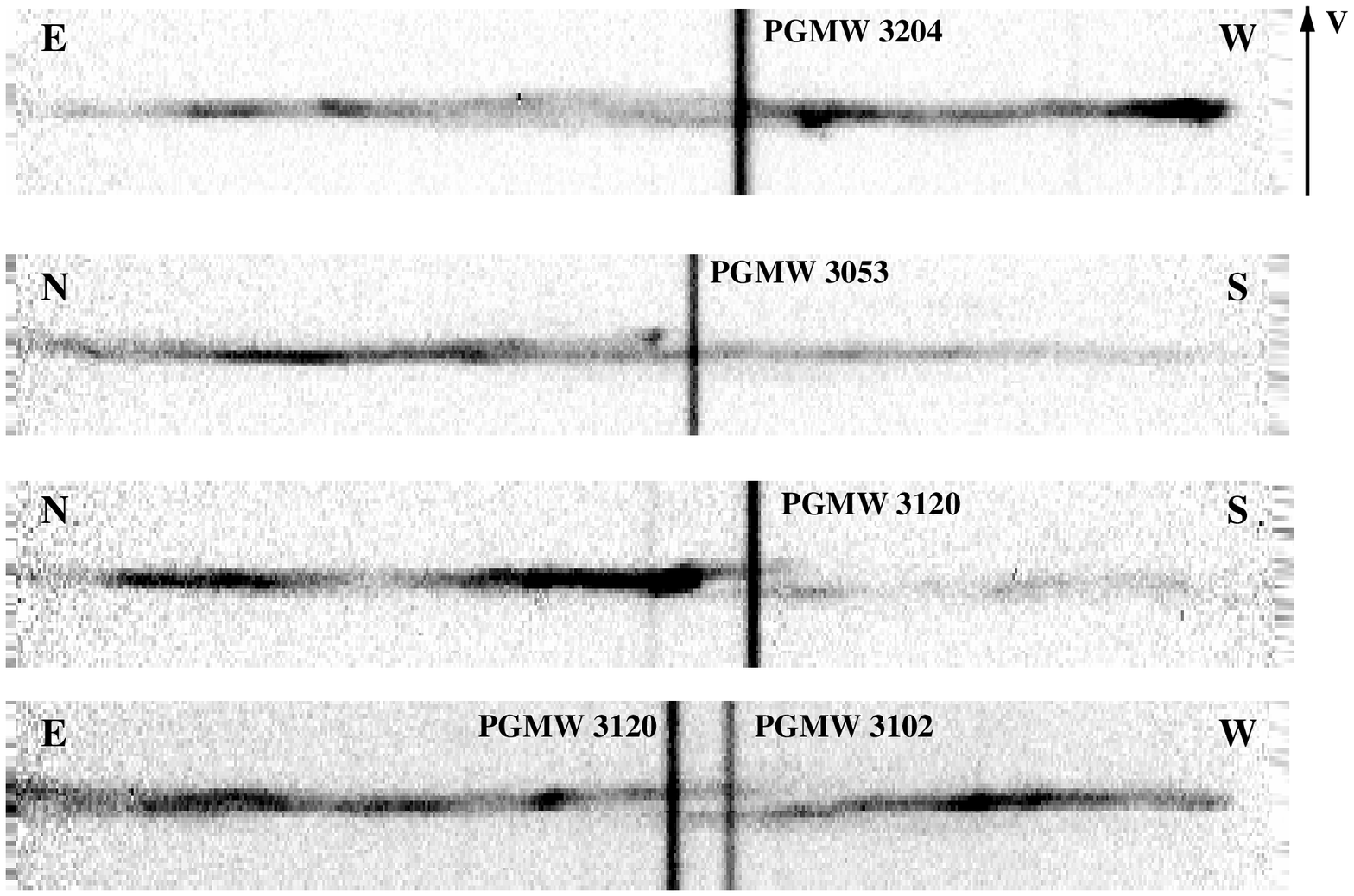}
 \includegraphics[width=0.4\textwidth]{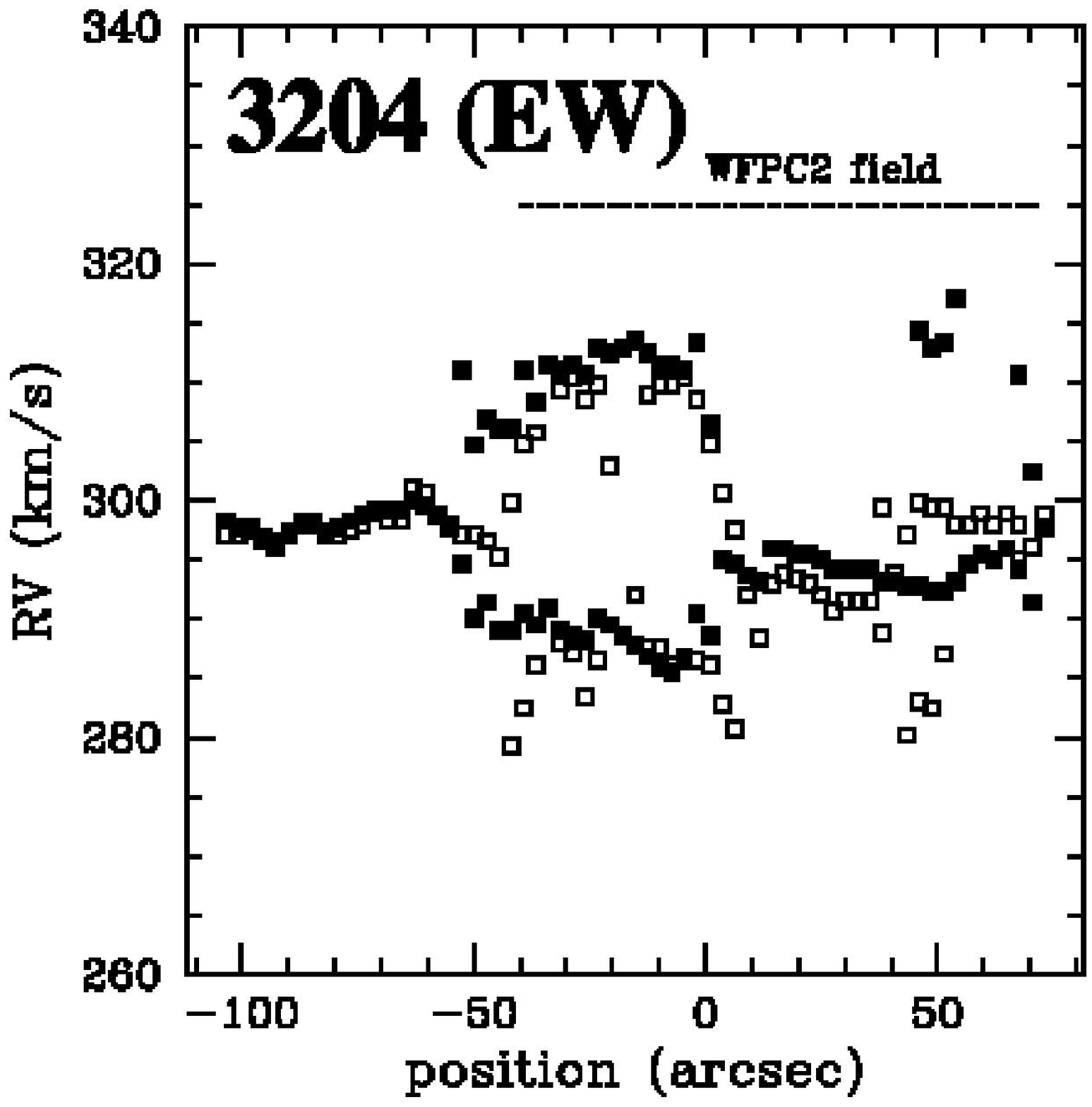}}
\caption{{\sl Top}: {\sl HST} WFPC2 H$\alpha$ image of N11B in 
the LMC, where 10$''$ corresponds to 2.5 pc. 
{\sl Bottom left}: Four echellograms of the [N\,II]$\lambda$6583
line in N11B.  Each panel is 185$''$ along the horizontal (spatial) 
axis, and 200 km s$^{-1}$ along the vertical (dispersion) direction.
{\sl Bottom Right}: The velocity-position plot along the E-W slit 
centered on the star PGMW-3204 (marked in the image above).
The echelle line image does not show obvious line-split, but the 
velocity components determined from line profile fitting show a 
clear velocity ellipse in both the H$\alpha$ (filled squares) and 
the [N\,II]$\lambda$6583 (open squares) lines.  This figure is 
from Naz\'e et al.\ (2001).}
\end{figure}

\vskip 6pt
\centerline{\bf Q1: Why Don't We See More Interstellar Bubbles?} 

Theoretically, every massive O star should be surrounded by an 
interstellar bubble, and as long the ambient interstellar gas
is dense enough the bubble should be visible as a ``ring nebula'' 
in H$\alpha$ images.
However, the Bubble Nebula (NGC\,7635) appears to be an exception,
rather than the rule.  Why don't we see more interstellar bubbles
around O stars?

The best way to answer this question is to observe H\,II regions of
young OB associations, where no supernova explosions have occurred
and the ambient interstellar gas is visibly dense.
This has been attempted by using {\sl HST} WFPC2 images of the H\,II
regions N11B and N180B in the Large Magellanic Cloud (LMC), but no
bubbles can be identified morphologically (Fig.~1)
The presence of interstellar bubbles in these young H\,II regions is 
nevertheless clearly illustrated by long-slit echelle spectroscopic
observations, in which expanding shell structures around early O 
stars are detected (Fig.~1). 
These expanding shells, centered on concentrations of O stars,
often exhibit different brightness in the approaching and receding 
sides, suggesting that the stars are formed on the surface of a 
cloud and their bubbles have a ``blister'' structure.

Why can't the expanding bubbles in N11B and N180B be morphologically 
identified in H$\alpha$ images?  While the bright background H\,II
region makes it difficult to detect the surface brightness 
perturbations caused by bubbles, the most important reason is that
the swept-up shell is not strongly compressed to raise its emission
measure.
The compression of a bubble shell depends on the shocks advancing 
into the ambient medium.  
As shown in Fig.~1, the expansion velocities of the bubbles in young
H\,II regions are only 10-15 km~s$^{-1}$, comparable to or slightly
higher than the isothermal sound velocity of the 10$^4$~K H\,II region,
$\sim$10 km~s$^{-1}$.
Therefore, the expanding bubbles hardly compress the ambient medium 
to produce the density contrast needed to exhibit a ring nebula 
morphology in H$\alpha$ images (Naz\'e et al.\ 2001).

Interestingly, H\,I 21-cm line observations frequently detect
interstellar bubbles around WR stars with expansion velocities
of $\sim$10 km~s$^{-1}$ (Cappa et al.\ 2003).
These interstellar bubbles are consistent with the bubbles seen in 
the young H\,II regions N11B and N180B.
As massive stars evolve and lose ionizing power, their interstellar
bubbles and ambient medium will recombine and cool.
The isothermal sound velocity of an H\,I medium is $\sim$1 km~s$^{-1}$,
so the expansion of the interstellar bubbles becomes highly supersonic
and causes stronger compression, producing larger contrast between the
bubble and the background.

\vskip 6pt
\centerline{\bf Q2: How Hot Is the Superbubble/Bubble Interior?}

The hot gas in bubble interiors can be detected in X-rays.
{\sl Einstein} and {\sl ROSAT} observations have revealed 
diffuse X-ray emission from two WR bubbles (NGC\,6888 and
S\,308), and a large number of superbubbles in the LMC.
It has been demonstrated that superbubbles in a quiescent
state are X-ray-faint with luminosities at least an order
of magnitude lower than expected from Weaver et al.'s (1977)
interstellar bubble models (Chu et al.\ 1995), and that 
superbubbles can be intermittently heated by supernovae near
the shell walls and become X-ray-bright (Chu \& Mac Low 1990).
The impact of supernovae not only generates bright X-ray
emission but also may produce breakouts through which hot
gas is vented to the surroundings, as shown in the LMC
superbubble N44 (Chu et al.\ 1993; Magnier et al.\ 1996).

\begin{figure}
\centerline{{\includegraphics[angle=-90,width=0.3\textwidth]{chu_f2a.eps}}
    \hskip 0.5cm 
       {\includegraphics[angle=-90,width=0.3\textwidth]{chu_f2b.eps}}}
\vskip -0.2cm
\caption{{\sl Chandra} ACIS-S spectra of the diffuse
 X-ray emission from the Omega Nebula (left) and the
 Rosette Nebula (right).  The observed spectra and best model
 fits are plotted in the top panels, and the residuals of the
 fits are in the bottom panels.  This figure is taken from
 Townsley et al.\ (2003).}

\vskip 0.3cm

\centerline{\includegraphics[width=0.6\textwidth]{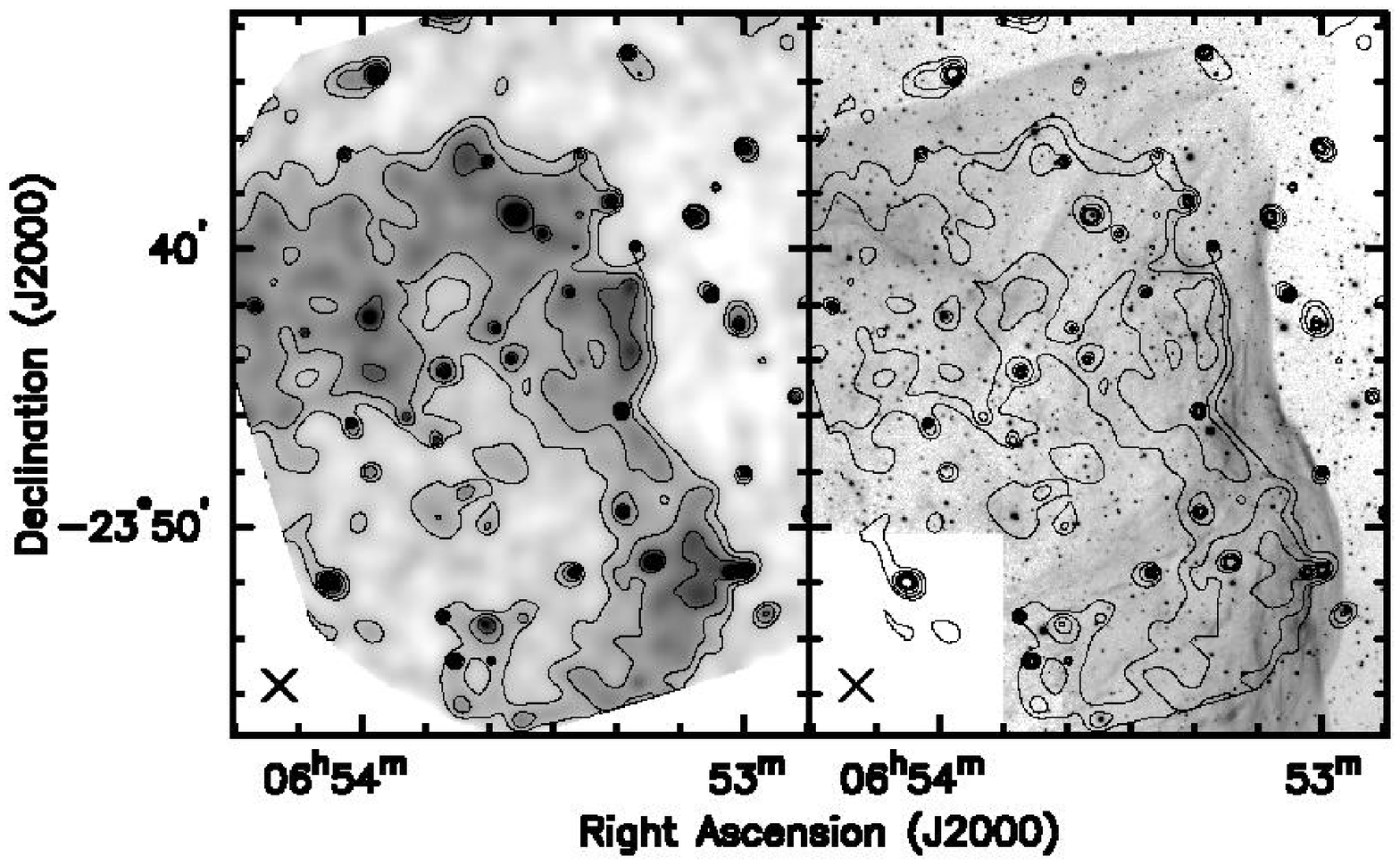}}
\vskip -0.2cm
\caption{{\sl Left}: {\sl XMM-Newton} EPIC X-ray image of the 
 NW quadrant of S\,308, and optical [O\,III]$\lambda$5007
 image of S\,308 overlaid by X-ray contours extracted from the EPIC
 image.  Note the distinct gap between the outer edges of X-ray
 emission and optical shell. The central WR star HD\,50896 is 
 marked with a``$\times$''.  This figure is adapted from Chu 
 et al.\ (2003).}

\vskip 0.3cm

\centerline{\includegraphics[width=0.7\textwidth]{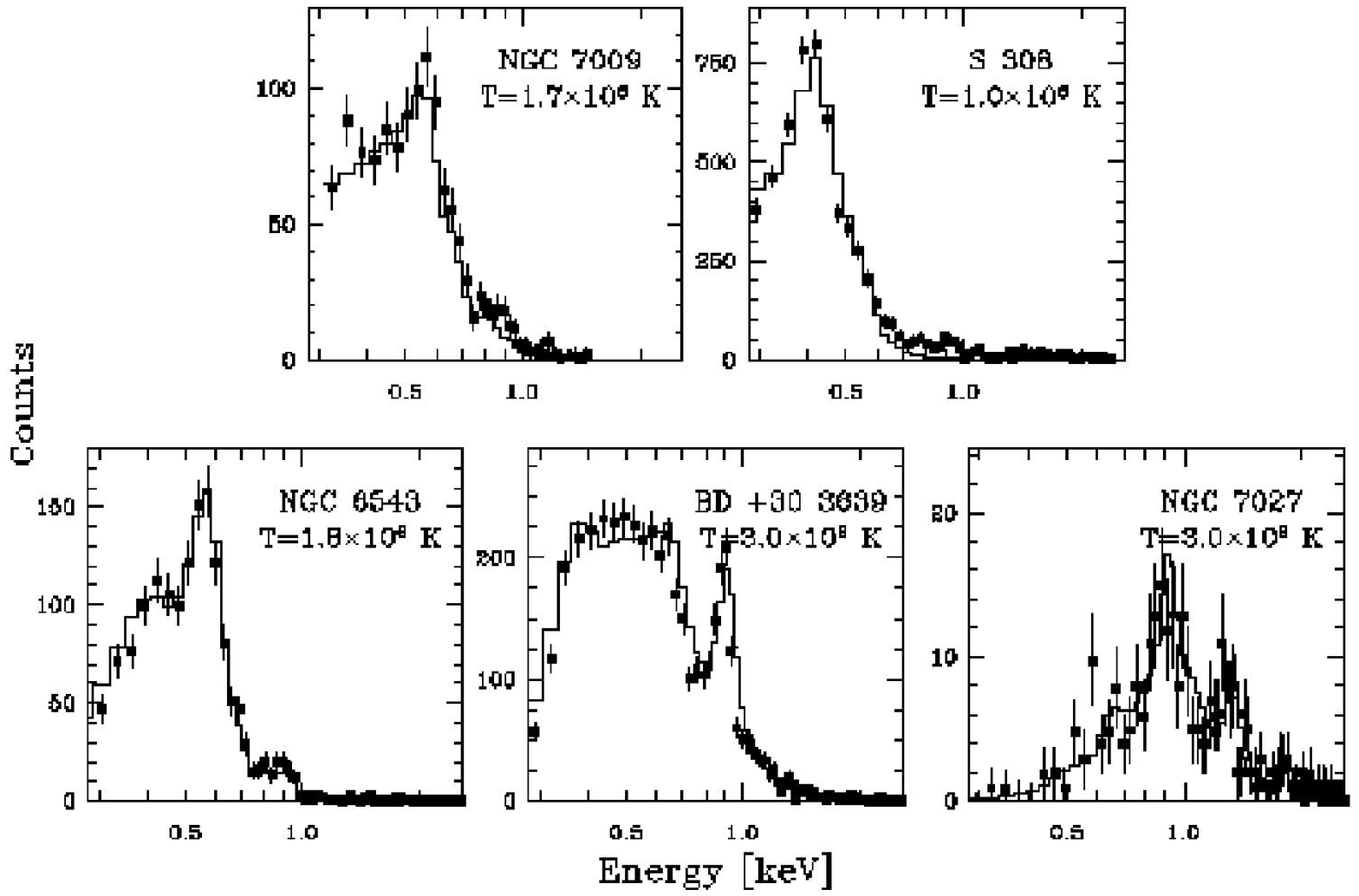}}
\vskip -0.2cm
\caption{{\sl XMM-Newton} EPIC spectra of the WR bubble S\,308
and PN NGC\,7009, and {\sl Chandra} ACIS-S spectra of the PNe
BD+30$^\circ$3639, NGC\,6543, and NGC\,7027.}
\end{figure}

{\sl Chandra} and {\sl XMM-Newton X-ray Observatories} have
made great strides in detecting hot gas in superbubble and
bubble interiors.
The most interesting superbubbles are the young ones that 
have not hosted any supernovae, so that their hot gas is
energized purely by fast stellar winds and can be compared
with the hot gas in single-star bubbles.
{\sl Chandra} has detected diffuse X-ray emission from two 
of such young superbubbles in the Galaxy, the Rosette Nebula 
and the Omega Nebula (Townsley et al.\ 2003).
Their X-ray spectra (Fig.~2) indicate that the hot gas has a 
major component at temperatures of almost 10$^7$ K and a minor 
component at temperatures of $\sim 1\times10^6$ K. 

No new detections of diffuse X-ray emission from single-star
bubbles have been made.  NGC\,6888 and S\,308 remain the only
two bubbles with known diffuse X-ray emission.
{\sl Chandra} ACIS-S observations of NGC\,6888 (Gruendl et al.\
2004) and {\sl XMM-Newton} EPIC observations of S\,308 (Chu et 
al.\ 2003) show differences in both the distribution and the
temperature of the interior hot gas.
The X-ray image of S\,308 shows a distinct gap between the outer
edge of the X-ray emission and the outer rim of the optical
shell (Fig.~3), while the diffuse X-ray emission from NGC\,6888 
reaches all the way to the rim of the optical shell.
The X-ray spectrum of diffuse emission from S\,308 (Fig.~4) is
extremely soft, indicating a plasma temperature of only 
$\sim1\times10^6$ K.
The X-ray spectrum of diffuse emission from NGC\,6888 is also 
soft, but the plasma temperature is slightly higher at 2--3 
$\times10^6$ K.

{\sl Chandra} and {\sl XMM-Newton} have detected shocked fast 
stellar winds from 5 PNe: BD+30$^\circ$3639, Mz-3, NGC\,6543, 
NGC\,7009, and NGC\,7027 (Kastner et al.\ 2000, 2001, 2003;
Chu et al.\ 2001; Guerrero et al.\ 2002).  
Their X-ray spectra, four shown in Fig.~4, are all soft, and 
their best spectral fits indicate plasma temperatures of 2--3
$\times10^6$ K.

It is interesting to compare the physical properties of hot 
gas among the superbubbles, WR bubbles, and PNe. 
In bubbles blown by a single star the hot gas temperatures are
all low, 1--3 $\times10^6$ K, while in superbubbles blown by large
numbers of massive stars the hot gas temperatures are much higher,
reaching 10$^7$ K.
The X-ray morphologies of single-star bubbles show 
limb-brightening, while those of young superbubbles show brighter
emission near stars at the center.
These results suggest that the hot gas in superbubbles is dominated
by colliding stellar winds in the vicinity of massive stars,
while the hot gas in single-star bubbles is generated by shocked
fast wind mixing with nebular material.
The mixing of cool nebular material into the hot gas is not simply
thermal evaporation, as the observed X-ray luminosities are much
lower than those expected from models with thermal conduction at 
the interface layer.
This problem leads to the next question.

\vskip 6pt
\centerline{\bf Q3: What Is Going on at the Hot/Cold Gas Interfaces in a
            Bubble?}

\begin{figure}
\centerline{{\includegraphics[width=0.3\textwidth]{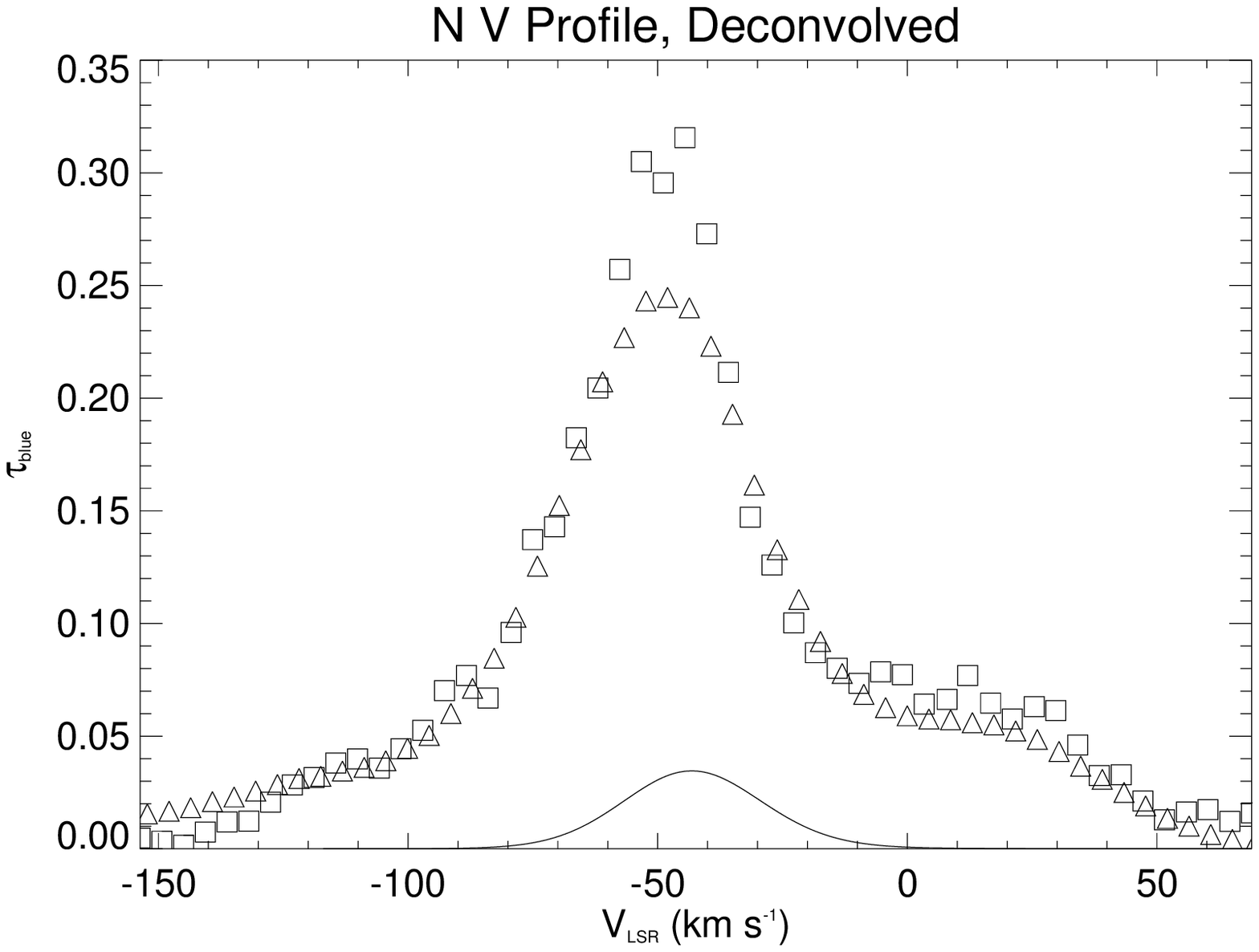}}
    \hskip 0.1cm 
            {\includegraphics[width=0.3\textwidth]{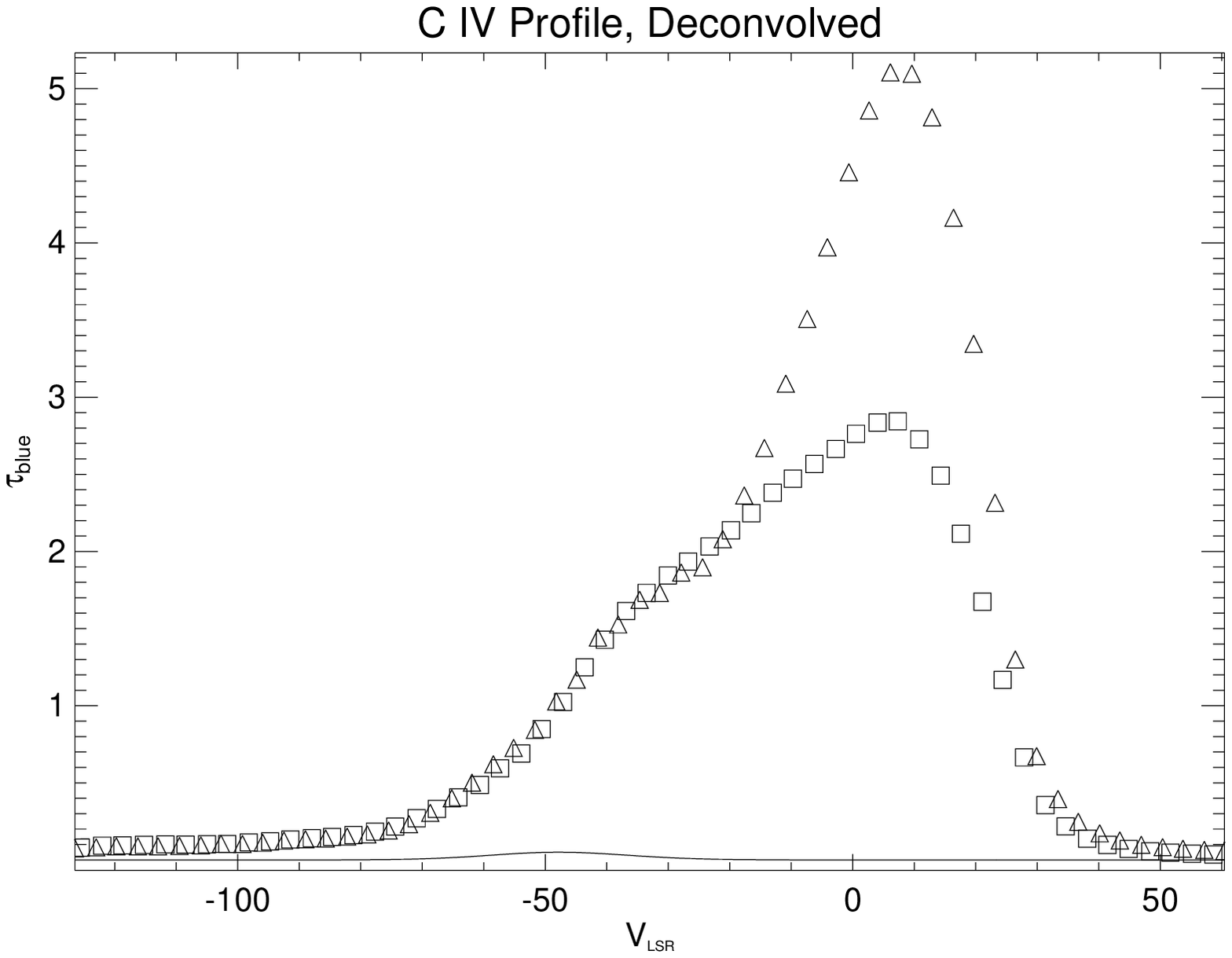}}
    \hskip 0.1cm 
            {\includegraphics[width=0.3\textwidth]{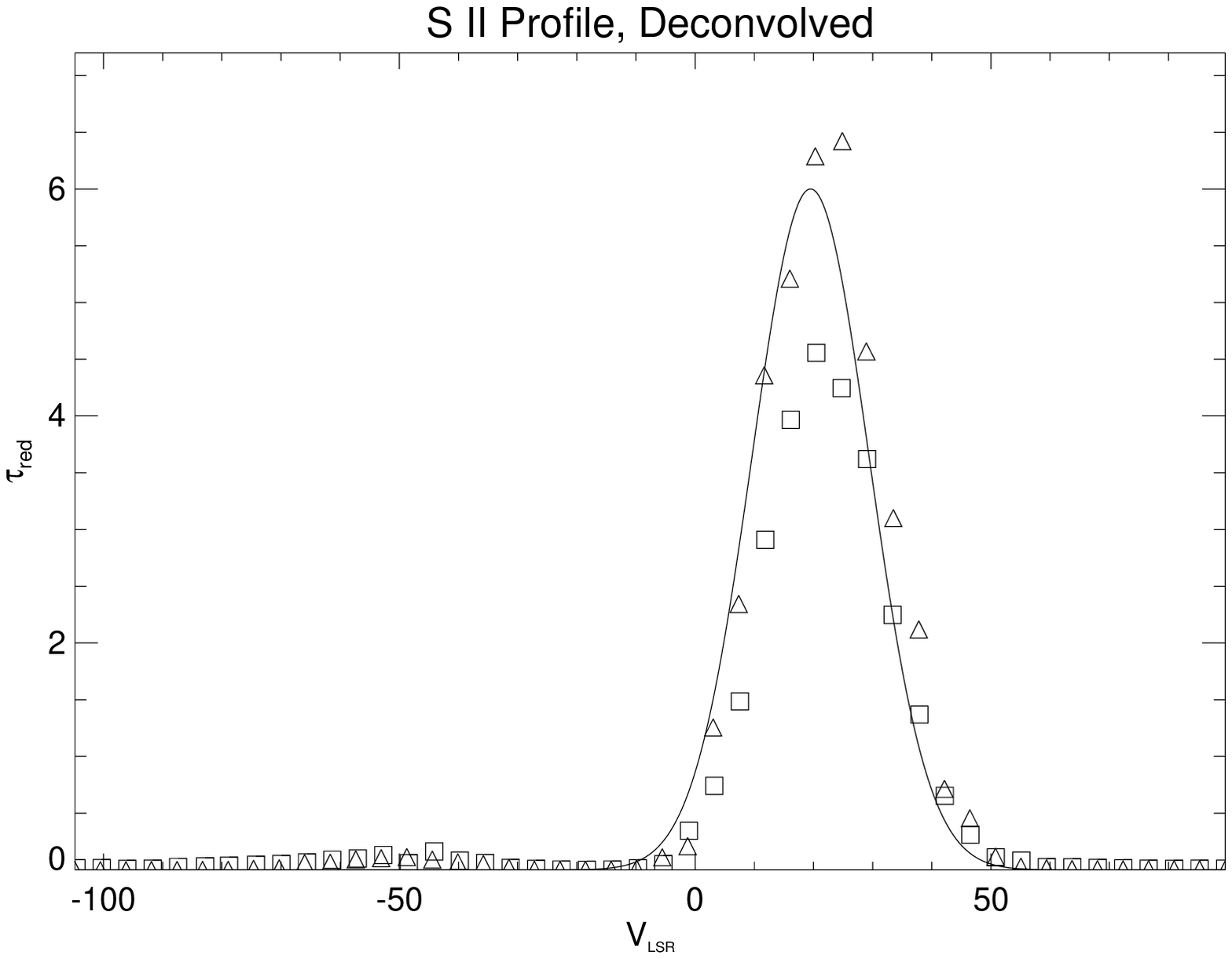}}}
\vskip -0.2cm
\caption{{\sl HST} GHRS observations of circumstellar/interstellar
  absorption toward HD\,50896, the central star of S\,308.  These 
  plots are optical depths of N\,V, C\,IV, and S\,II lines as 
  functions of V$_{\rm LSR}$, taken from Boroson et al.\ (1997).}
\end{figure}

The interfaces are conventionally observed with absorption lines of
highly ionized species such as C\,IV and N\,V.
The first convincing detection of a single interface was reported by
Boroson et al.\ (1997) in the WR bubble S\,308 (Fig.~5);
the $-$50 km~s$^{-1}$ component of the N\,V line is identified as 
originating from the interface on the approaching side of the bubble.
It is evident that the S\,II absorption originates from low-ionization
gas along the line of sight, and that C\,IV, peaking at the same 
velocity, is most likely photoionized unlike N\,V which is 
collisionally ionized.

The {\sl Far UV Spectroscopic Explorer (FUSE)} opened up a new window 
to observe the interface -- the O\,VI lines.
The two WR bubbles with known diffuse X-ray emission provide the best 
laboratory to study interfaces.
{\sl FUSE} observations of the O\,VI emission and {\sl HST}
long-slit STIS observations of the N\,V emission may spatially
resolve the temperature structure of the interfaces.
It will then be possible to assess the relative importance of 
thermal evaporation and dynamical ablation for mixing nebular
material into the hot bubble interior.
We do not have an answer to the question on the interface yet, but
we are working on it and will find the truth one day.

\section{Conclusion}

{\sl HST}, {\sl Chandra}, {\sl XMM-Newton}, and {\sl FUSE} observations
of bubbles and superbubbles are beginning to reveal their physical 
structure.
These results will provide valuable guidance for future theoretical 
modeling of bubbles.

\vskip -0.5cm
\noindent{\bf\large 4. ~~~~~~Discussion} \\

\noindent
{\it Welsh:} The trouble with using Weaver et al. equation 
is that there is a huge ``fudge factor'' called $n_0$, the
ambient interstellar density which can be ``tweaked'' to
give almost any answer you want. \\

\noindent
{\it Chu:} It is possible to use the surface brightness of
the ambient H\,II region to estimate the ambient density,
or imply the ambient density from the swept-up shell density.
See the analysis of bubbles in  N11B and N180B by Naz\'e 
et al.\ (2001), and the analysis of the superbubble in M17 
by Dunne et al.\ (2003). \\

\noindent
{\it Hester:} The only H$\alpha$ spectrum you showed had two 
components that did not join up into a velocity ellipse.
How do you argue that this is a bubble rather than a couple of 
different evaporative outflows along the same line of sight? \\

\noindent
{\it Chu:} The [N II]$\lambda$6583 echellograms I showed 
were not well displayed to show the faint emission.  Most split 
line profiles have one component stronger than the other, 
indicating a ``blister'' structure.  The velocity ellipses are 
closed.  The evaporative outflows would start from the systemic 
velocity and extend to blue-shifted or red-shifted velocities 
continuously.  This is not supported by the velocity ellipses 
we observe, as there are no stationary components at the
systemic velocity in the velocity ellipses. \\

\noindent
{\it Ma\'{\i}z-Apell\'aniz:} One of the reasons why it is hard to
see bubbles morphologically in a region like N11B is that most of
the optical nebular photons originate on the H\,II region itself 
(at the interface with the molecular cloud).  It will be difficult
to distinguish dim bubbles against the bright background. \\

\noindent
{\it Chu:} See answer in the text.

\end{document}